\documentclass{PoS}
\usepackage{amsmath,amssymb}
\usepackage{graphicx}
\newcommand{\Comments}[1]{}

\newcommand{\nn}{\nonumber}

\newcommand{\chibar}{{\bar{\chi}}}
\newcommand{\Psfig}[2]{\includegraphics[width=#1]{Figs/#2}}
\newcommand{\Feff}[2]{{\cal F}_\mathrm{#1}^{#2}}

\newcommand{\Fq}{{\cal F}_\mathrm{q}}

\newcommand{\Fint}{{\cal D}}
\newcommand{\Seff}[1]{S_\mathrm{eff}^{(#1)}}
\newcommand{\comment}[1]{}

\newcommand{\Zchi}{Z_{\chi}}

\newcommand{\tilmu}{\tilde{\mu}}

\newcommand{\bp}{\beta_{p}}

\newcommand{\lbar}{\bar{\ell}}

\newcommand{\bfx}{{\bf x}}

\title{Effective Potential and Phase Diagram \\
in the Strong-Coupling Lattice QCD \\ with Next-to-Next-to-Leading Order \\ and Polyakov Loop Effects}

\ShortTitle{Phase Diagram in the Strong-Coupling Lattice QCD with Polyakov Loop Effects}

\author{\speaker{Takashi Z. Nakano} \\ %\thanks{A footnote may follow.}\\
        Department of Physics, Faculty of Science, Kyoto University,
Kyoto 606-8502, Japan \\
        Yukawa Institute for Theoretical Physics, Kyoto University,
Kyoto 606-8502, Japan \\
        E-mail: \email{tnakano@yukawa.kyoto-u.ac.jp}}

\author{Kohtaroh Miura\\
        INFN-Laboratori Nazionali di Frascati, I-00044, Frascati(RM), Italy\\
        }

\author{Akira Ohnishi\\
        Yukawa Institute for Theoretical Physics, Kyoto University,
Kyoto 606-8502, Japan \\
        }

\abstract{We investigate chiral and deconfinement transitions
in the strong coupling lattice QCD for color SU(3).
We combine the leading order Polyakov loop effective action
of the strong coupling expansion
and the next-to-next-to-leading order ($1/g^4$) fermionic effective action
with one species of unrooted staggered fermion.
Two approximation schemes are adopted to evaluate the Polyakov loop effects; 
a Haar measure method (no fluctuation from the mean field)
and a Weiss mean-field method (with fluctuations).
The Polyakov loop is found to suppress the chiral condensate
and to reduce the chiral transition temperature at $\mu=0$.
The chiral transition temperature roughly reproduces the Monte Carlo results
in the region
$\beta=2N_c/g^2 \lesssim 4$.
}

\FullConference{The XXVIII International Symposium on Lattice Field Theory, Lattice2010\\
		June 14-19, 2010\\
		Villasimius, Italy}

\begin{document}

\section{Introduction}
The phase transition in Quantum Chromodynamics (QCD) 
at finite temperature ($T$) and/or quark chemical potential ($\mu$)
is attracting much attention in recent years.
Since lattice Monte Carlo simulations have the notorious sign problem
at finite $\mu$~\cite{MC-sign-problem}, we need to invoke some approximations in QCD
or effective models at finite chemical potential.
Strong coupling lattice QCD (SC-LQCD)
is one of the most promising approximation schemes in QCD to investigate the chiral and deconfinement 
phase transitions at finite $T$ and $\mu$.
The QCD phase transition has two characteristic features;
the restoration of the chiral symmetry which is spontaneously broken
in vacuum (chiral transition),
and the liberation of color degrees of freedom which is confined
at low temperatures (deconfinement transition).
The chiral transition at finite $T$ and $\mu$ has been investigated in SC-LQCD~\cite{DKS,finiteTtreatment,SC-QCDphasediagram,NLONNLOSC-LQCD}.
The deconfinement transition at finite $T$ was qualitatively explained
on the basis of the mean-field treatment of the Polyakov loop
in the leading order of the strong coupling expansion~\cite{Kogut:1981ez}.
Higher order corrections on the Polyakov loop action has been
investigated recently~\cite{Langelage}.
One of the interesting developments describing the QCD phase transition
at finite $T$ and $\mu$
can be found in the works~\cite{GOIKmodel,FukushimaGOmodel},
which include
the leading order Polyakov loop effective action 
(finite coupling constant) and
the strong coupling limit effective action for quarks (infinite coupling constant),
and enables us to describe the chiral and deconfinement transition
in a single framework.
To make a step forward towards the true phase diagram,
it is necessary to consider Polyakov loop effects and finite coupling effects with quarks.

In this proceedings,
we develop a {\em Polyakov loop extended strong coupling lattice QCD (P-SC-LQCD)}
framework~\cite{Nakano:2010bg} by combining
the leading order Polyakov loop effective action
and NNLO quark effective action.
We derive an analytic expression of the effective potential
in P-SC-LQCD at finite $T$ and $\mu$,
and investigate the chiral and deconfinement phase transitions at $\mu=0$.

\section{Effective Potential with Polyakov Loop effects}\label{sec:Feff}

In the lattice QCD, the partition function and action
with one species of unrooted staggered fermion
for color $\mathit{SU}(N_c)$ 
in the Euclidean spacetime
are given as,
\begin{align}
\label{Eq:ZLQCD}
{\cal Z}_{\mathrm{LQCD}} 
=& \int \Fint[\chi,\chibar,U_\nu]~e^{-S_\mathrm{LQCD}}
\ ,\quad %\\
S_\mathrm{LQCD}=S_F+S_G
\ ,\\
S_F
=&\frac {1}{2} \sum_x \sum_{\rho=0}^d 
  \left[ \eta_{\rho,x} \bar{\chi}_x U_{\rho,x} \chi_{x + \hat{\rho}} 
  - \eta_{\rho,x}^{-1} \left( h.c. \right) \right]
%  
%  \bar{\chi}_{x + \hat{\rho}}
%   U_{\rho,x}^{\dagger} \chi_x \right]
%\nn\\
+m_0 \sum_x \bar{\chi}_x \chi_x 
%\ ,\\
\ , \quad 
S_G
 =%&
  -\frac {2}{g^2} \sum_{P} 
 	\mathrm{Re} U_{P},
%\frac {2N_c}{g^2} \sum_{P} \left[
% 	1 - \displaystyle \frac{1}{2N_c} \left( U_P + U_P^\dagger \right) \right] ,
\end{align}
where
$\chi (\bar{\chi})$, $m_0$, $U_{\rho,x}$,
and
$U_P$ 
denote
the quark (antiquark) field,
bare quark mass, link variable,
and plaquette, respectively.
In the staggered fermion,
the spinor index is reduced to the staggered phase factor 
$\eta_{\rho,x}=(\eta_{0,x},\eta_{j,x})=(e^\mu, (-1)^{x_0+\cdots +x_{j-1}})$ where $\mu$ is the quark chemical potential.
Note that we utilize the lattice unit $a=1$
and represents physical values as dimensionless values normalized
by the lattice spacing $a$ throughout this proceedings.

To treat the chiral and deconfinement phase transitions simultaneously,
we evaluate the leading order Polyakov loop effects
in the pure Yang-Mills sector
and NNLO effects in the fermionic sector.
In a finite $T$ treatment of SC-LQCD,
we first derive an effective action
by carrying out spatial link ($U_j$) integrals,
and evaluate the temporal link ($U_0$) integral later to consider
the thermal effects of quarks~\cite{finiteTtreatment}.

\begin{figure}[th]
\begin{center}
\Psfig{8cm}{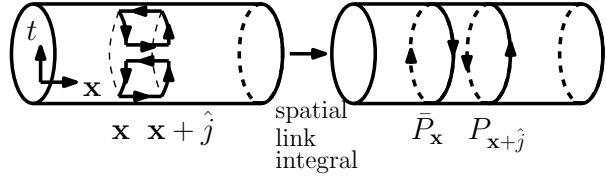}
\caption{
Leading order of the Polyakov loop effects in the strong coupling expansion.
The squares in the left and loops in the right 
represent the temporal plaquettes 
and the Polyakov loops, respectively.}
\label{Fig:Polyakov-loop-SC-LQCD}
\end{center}
\end{figure}

We can obtain the leading order Polyakov loop effective action $\Seff{P}$
from the spatial link integral
of the $N_\tau$ plaquettes sequentially connected in the temporal direction
as shown in Fig.~\ref{Fig:Polyakov-loop-SC-LQCD}.
\begin{align}
 \Seff{P}
=& -\left( \displaystyle \frac {1}{g^2 N_c} \right)^{N_{\tau}} N_c^2
     \sum_{\bfx, j>0} 
     \left(
      \bar{P}_\bfx P_{\bfx+ \hat{j}} + h.c.
     \right)
\ ,
\label{Eq:PLaction}
\end{align}
where $N_\tau$ and $P_\bfx = \mathrm{tr}_c \prod_\tau U_0 (\tau,{\bf x}) / N_c$
represent the temporal lattice size and Polyakov loop, respectively. 
This effective action shows the nearest-neighbor interaction
of the Polyakov loop.
The factor $1/g^2N_c$ arises from the spatial link integral.
By comparison, we derive NLO and NNLO quark effective action 
by the cluster expansion in Ref.~\cite{NLONNLOSC-LQCD}. 
% 
%
%
%
%============================================================
%============================================================
%
%
%
%
%
%
%
%
From a naive counting of the strong coupling expansion order,
the leading order Polyakov loop action composed of the plaquettes
is in the higher order ($\mathcal{O}(1/g^{2N_\tau})$), 
compared with the NNLO terms ($\mathcal{O}(1/g^{4})$)
stem from the quark sector.
To describe the chiral and deconfinement phase transitions,
we set the starting point of chiral and deconfinement dynamics
as the effective action of ${\cal O}(1/g^0)$ and ${\cal O}(1/g^{2N_\tau})$
in the quark and Polyakov loop sectors, respectively.
The Polyakov and NNLO quark effective action corresponds
to an extension in the quark sector
from this starting point.

From the effective action described above,
we obtain an approximate QCD partition function
and define the effective potential $\Feff{eff}{}$ as 
\begin{align}
\Feff{eff}{}
\equiv& -\frac{1}{N_\tau L^d}\log {\cal Z}_{\mathrm{LQCD}} 
%\nn\\
\simeq -\frac{1}{N_\tau L^d}\log\left[\int {\cal D}[\chi,\chibar,U_0]
	e^{-S_\mathrm{NNLO}-\Seff{P}}
	\right]
\nn\\
=&\Fq(\Phi;\mu,T)
 + U_g(\ell,\lbar)
 +\Feff{eff}{(X)}(\Phi)
\ ,
\end{align}
where $L$ and $d(=3)$ are the spatial lattice size and the spatial dimension, respectively.
Here $S_\mathrm{NNLO}, \Fq,U_g$,
and $\Feff{eff}{(X)}$
represent the NNLO quark effective action, quark free energy, pure gluonic potential,
and effective potential 
including only the auxiliary fields $\Phi$,
respectively.
In this work,
we first reduce the effective action 
to the bilinear form for the quark fields
by introducing several auxiliary fields,
and assume the mean-field values for $\Phi$ later.
We obtain $\Fq$ and $U_g$
by evaluating the Grassmann $(\chi,\bar{\chi})$
and temporal link ($U_0$) integrals,
In this proceedings, we evaluate the temporal link integral in two kinds of methods,
the Haar measure and Weiss mean-field methods.

We shall now derive the effective potential with Polyakov loop effects
in the Haar measure method (H-method).
In the H-method, we replace the Polyakov loop with a mean-field value
and take into account the Haar measure in the Polyakov loop potential
instead of carrying out the temporal link integral.
The contribution to the effective action is,
\begin{align}
 \Seff{P}
\simeq& -2 \bp L^d \lbar \ell 
\ , \label{Eq:PLactionHaar}
\end{align}
where $\beta_p = (1/g^2 N_c)^{N_\tau} N_c^2 d$,
$\ell=\langle P_\bfx \rangle$ and $\lbar=\langle \bar{P}_\bfx \rangle$.
We assume the mean fields $\ell$ and $\lbar$
are constant and isotropic.
The temporal link integral is represented by using
the Haar measure
in the Polyakov gauge~\cite{DKS}, which is a static and diagonalized gauge 
for temporal link variables,
\begin{align}
\int d\mathcal{U}_0
=& \int d \ell d\lbar \cdot 27 
\left[ 1 -6 \ell \lbar +4 \left( \ell^3 + \lbar^3 \right) - 3 \left( \ell \lbar \right)^2 
\right]
\ , 
\end{align}
where ${\cal U}_0(\bold{x})=\prod_\tau U_0(\bold{x},\tau)$. 
This Haar measure shows the Jacobian in the transformation
from the temporal link variables ($U_0$)
to the Polyakov loop ($\ell,\lbar$).
$\Fq$ and $U_g$ are given as,
\begin{align}
&\Fq
=
- N_c E_q
- T \log R(E_q-\tilmu,N_c\ell, N_c\lbar)
%\nn\\
%&~~~~~~~~~
- T \log R(E_q+\tilmu,N_c\lbar, N_c\ell)
- N_c \log \Zchi
\ ,\label{Eq:FqHaar}\\
&R(x,L,\bar{L})\equiv
1+Le^{-x/T}+\bar{L}e^{-2x/T}+e^{-3x/T}
\ , \\
&U_g
=
-2 T \bp \lbar \ell 
-T \log \left[
1 -6 \ell \lbar +4 \left( \ell^3 + \lbar^3 \right) - 3 \left( \ell \lbar \right)^2 
\right]
\ ,
\label{Eq:UgHaar}
\end{align}
where $\beta_p = (1/g^2 N_c)^{1/T} N_c^2 d$,
and $E_q$ is the quark excitation energy.
Here we have replaced the $N_\tau$ with $1/T$ and
omitted irrelevant constants.
$\Fq$ includes the vacuum, quark, and antiquark 
free energies and the contribution of the wave function renormalization factor $\Zchi$.
The quark free energy includes one- and two-quark excitations $(e^{-(E-\tilmu)/T},e^{-2(E-\tilmu)/T})$.
In the confined phase ($\ell \sim \lbar \sim 0$), 
the one- and two- quark excitations are suppressed and
only the color-singlet state contributions remain.
The pure gluonic potential $U_g$ does not include the fluctuation of the Polyakov loop since we 
treat the Polyakov loop as the mean field without the temporal link integral.
In the Polyakov-loop extended Nambu-Jona-Lasino model, this pure gluonic potential is incorporated
to express the properties of the deconfinement phase transition~\cite{Fukushima:2003fw}.

Now we shall evaluate the Polyakov loop effects
in the Weiss mean-field method (W-method).
The W-method 
includes some part of the fluctuation effects of the Polyakov loop.
We first bosonize the Polyakov loop action
by using the {\it Extended Hubbard-Stratonovich (EHS) transformation}~\cite{NLONNLOSC-LQCD},
which is a procedure to bosonize the product
of different types of composites.
Then, the Polyakov loop action in Eq.~(\ref{Eq:PLaction}) is linearized as,
\begin{align}
 \Seff{P}
\approx& \left( \displaystyle \frac {1}{g^2 N_c} \right)^{N_{\tau}} N_c^2  \sum_{{\bf x}, j>0} 2
     \left(
      \lbar \ell - \bar{P}_\bfx \ell - \lbar P_\bfx
     \right) %\nonumber\\
\simeq  2 \bp L^d \lbar \ell 
 -2 \beta_p \sum_{\bf x} \left( \bar{P}_\bfx \ell + \lbar P_\bfx \right)
\ ,
\label{Eq:PLactionEHS}
\end{align}
where $\bp,P_\bfx$, and $\bar{P}_\bfx$ are defined before.
$\ell$ and $\lbar$ represent the auxiliary fields for the Polyakov loop,
$(\ell=\langle P_\bfx \rangle, \lbar=\langle \bar{P}_\bfx \rangle)$.
In ''$\simeq$'' of Eq. (\ref{Eq:PLactionEHS}),
we assume constant and isotropic values for auxiliary fields $\ell$ and $\lbar$.
We obtain $\Fq$ and $U_g$
by evaluating the Grassmann $(\chi,\bar{\chi})$
and temporal link ($U_0$) integrals,
\begin{align}
{\cal F}_q
= & %-N_c E_q
-T \log(Z_P/L_0) - N_c \log Z_\chi
\label{Eq:Fqboso}
\ , \\
Z_P
=&\int d\mathcal{U}_0\, \mathrm{det}_c \Bigl[
	2\cosh(E_q/T)
	+\mathcal{U}_0 e^{\tilmu/T}
	+\mathcal{U}_0^{\dagger} e^{-\tilmu/T}
	\Bigr]
%\nn\\
\exp
\left[\eta\mathrm{tr}\,{\cal U}_0^\dagger+\bar{\eta}\mathrm{tr}\,{\cal U}_0\right]
\label{Eq:eVqB}
\ , \\
U_g=& 2 T \bp \lbar \ell -T \log L_0
\label{Eq:Ugboso}
\ . 
\end{align}
We have defined $\eta=2\bp \ell /N_c$ and $\bar{\eta}=2\bp \lbar /N_c$.
In Eq. (\ref{Eq:eVqB}), we can perform the temporal link integral and obtain the analytic expression.
$Z_P$ and $L_0$ are functions of the combination of the modified Bessel functions.
In Ref.~\cite{Nakano:2010bg}, we show the explicit expression of them.
Compared with the effective potential in the H-method
where the quark contributions are represented as the vacuum, quark and antiquark parts
, Eq. (\ref{Eq:FqHaar}),
the effective potential is more complicated.

Note that the pure gluonic potential $U_g$ includes the dependence
on $\lbar/\ell$ explicitly (i.e. the dependence on $\mu$).
When the quark chemical potential is zero ($\mu=0$),
the Polyakov loop for quarks and antiquarks are the same $(\ell=\lbar)$.
In comparison,
when the quark chemical potential is finite ($\mu \neq 0$),
the Polyakov loop for antiquarks
is generally different from that for quarks
$(\ell \neq \lbar)$~\cite{llbarldiff}.

\section{Chiral and Deconfinement Phase Transitions}\label{sec:chiraldeconf}

\begin{figure}[htbp]
\begin{tabular}{cc}
\begin{minipage}{0.5\hsize}
\begin{center}
\Psfig{\hsize}{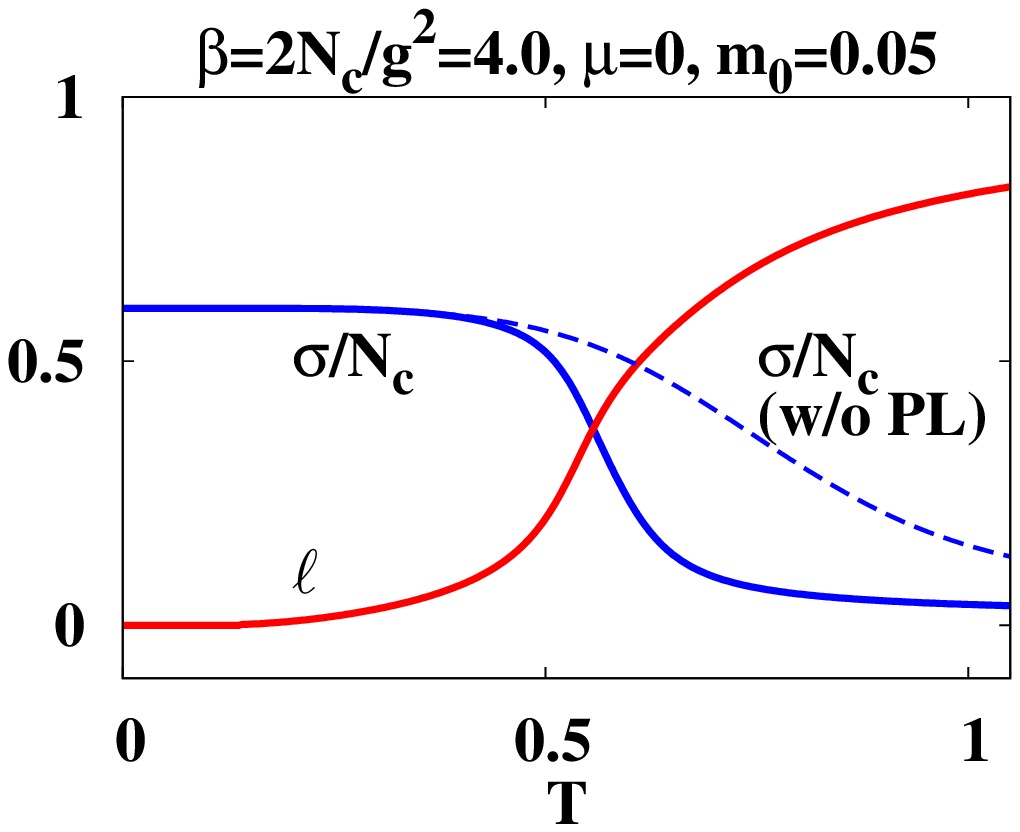}
\end{center}
\end{minipage}
\begin{minipage}{0.5\hsize}
\begin{center}
\Psfig{\hsize}{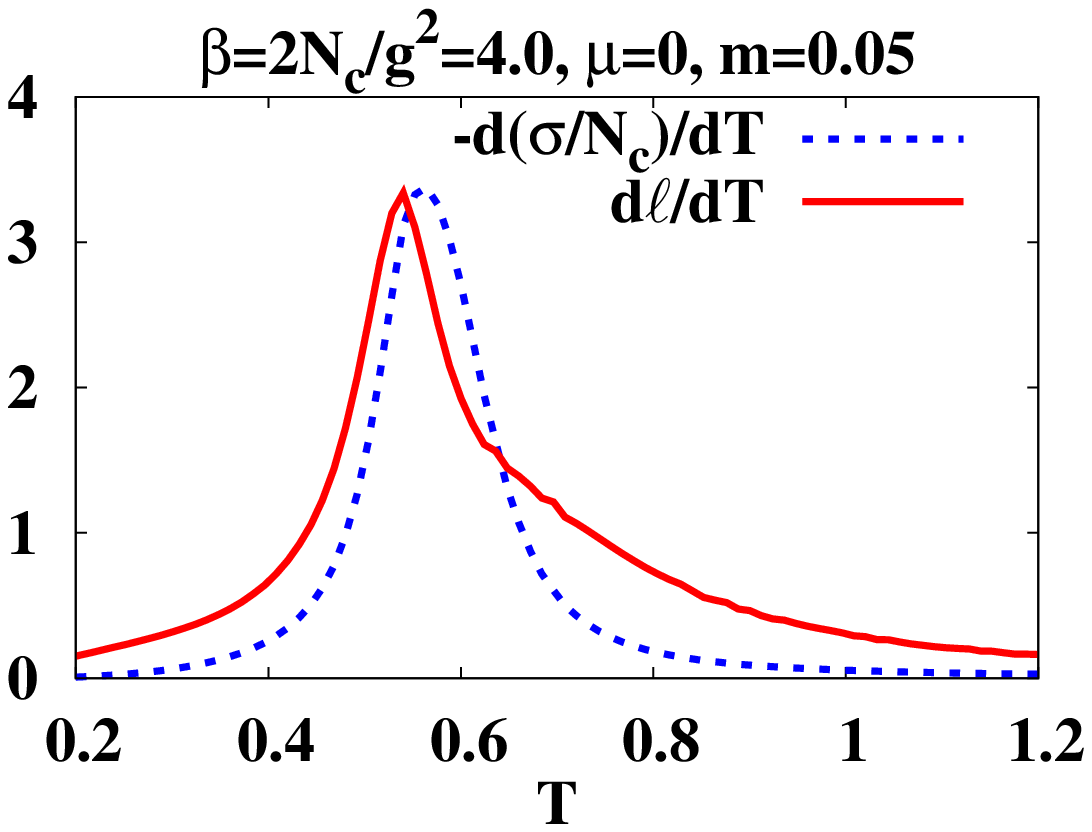}
\end{center}
\end{minipage}
\end{tabular}
\caption{
Left panel:
Chiral condensate and Polyakov loop 
in P-SC-LQCD (solid lines),
and chiral condensate in SC-LQCD without the Polyakov loop effects
(dashed line) as functions of $T$ at $\mu=0$ in the W-method.
Right panel: 
Temperature dependence of
$d\sigma/dT$ (dashed line) and $d \ell /dT$ (solid line) in the W-method.
}
\label{Fig:P-sig-PL}
\end{figure}

\begin{figure*}[tbh]
\begin{center}
\Psfig{8cm}{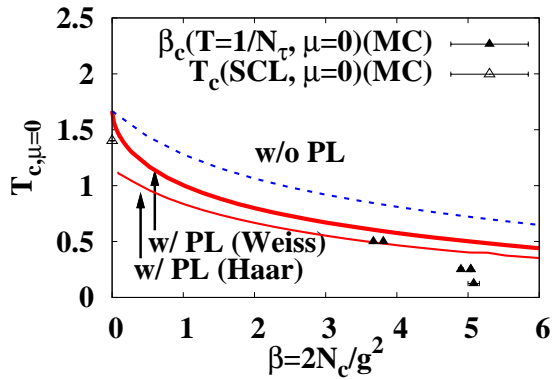}
\caption{
Comparison of chiral transition temperature between P-SC-LQCD (solid line) 
and SC-LQCD without the Polyakov loop (dashed line).
We show the results of Weiss mean-filed method (bolid solid line) and Haar measure method (thin solid line) in P-SC-LQCD.
We define the critical temperature as the peak of
$-d\sigma/dT$.
The triangles represent the results of
the critical temperature ($T_{c,\mu=0}$, open triangle)
and the critical coupling ($\beta_c$, filled triangles)
obtained in Monte-Carlo simulations
with one species of unrooted staggered fermion~\cite{deForcrand:2009dh,MCresults} 
}
\label{Fig:Tc-W-H}
\end{center}
\end{figure*}

In Fig. \ref{Fig:P-sig-PL},
we show the chiral condensate ($\sigma$),
Polyakov loop,
$d\sigma/dT$ and $d\ell/dT$ as functions of $T$.
The chiral and deconfinement phase transitions seems
to take place at almost the same $T$.

We find that Polyakov loop suppresses the chiral condensate
and reduces the chiral transition temperature in P-SC-LQCD.
In left panel of Fig.~\ref{Fig:P-sig-PL}
and Fig.~\ref{Fig:Tc-W-H},
we compare the chiral condensate $\sigma$
and the chiral transition temperature $T_{\chi,c}$
with and without Polyakov loop effects.
We find that both $\sigma$ and $T_{\chi,c}$ in P-SC-LQCD
are smaller than those without Polyakov loop effects;
the chiral condensate becomes smaller when the Polyakov loop
takes a finite value,
and the chiral symmetry restoration takes place at lower $T$
by the Polyakov loop effects.
In SC-LQCD without Polyakov loop effects,
we find the contribution only from 
color-singlet states,
then we implicitly assume that the quarks are confined at any $T$.
In the W-method,
we have one- and two-quark contribution as well
when the Polyakov loop takes a finite value.
Quark excitation generally breaks the chiral condensates,
then it promotes the chiral symmetry to be restored at lower $T$.
A similar behavior is found also in the H-method.
While qualitative behaviors are the same in both of the methods,
the W-method exhibits a little larger $T_{c}$ than the H-method.
Physically, the temporal link integral in the Weiss mean-field method favors
the color-singlet states and therefore suppress the quark excitation.

In our previous work on NLO
and NNLO SC-LQCD~\cite{NLONNLOSC-LQCD}, 
the critical temperature is calculated to be larger than MC results,
and NNLO effects on $T_{\chi,c}$ are found to be small.
In Fig. \ref{Fig:Tc-W-H}, 
we show $T_{\chi,c}$ in two treatments of P-SC-LQCD
in comparison with the MC results.
The MC results, especially in the region $\beta \lesssim 4$,
are roughly explained
in P-SC-LQCD developed in this work.
Namely, the Polyakov loop and and finite coupling (NLO and NNLO) effects
dominantly contribute to the reduction of $T_{\chi,c}$.
This observation implies that introducing the Polyakov loop,
the deconfinement order parameter, is essential to explain 
the QCD phase transition temperature.

\section{Concluding Remarks}\label{sec:CR}
In this proceedings,
we have derived an analytic expression of the effective potential
at finite temperature and chemical potential
in the strong coupling lattice QCD with the Polaykov loop effects
using one species of unrooted staggered fermion.
The chiral and deconfinement transitions at $\mu=0$ are discussed with emphasis on
the Polyakov loop effects on these transitions.
The NNLO quark effective action in the strong coupling expansion
is combined with the leading order 
Polyakov loop action,
${\cal O}(1/g^{2N_\tau})$,
and we have evaluated the temporal link ($U_0$) integral in two methods.
One is the Haar measure method (H-method),
where we replace the Polyakov loop
with its constant mean-field without the $U_0$ integral.
The deconfinement dynamics is taken into account via
the Haar measure.
Another is the Weiss mean-field method (W-method),
where we bosonize the effective action  
and carry out the temporal link integral explicitly,
then the fluctuation effects are included.
We have found that one- and two-quark excitations are allowed in both methods,
when the Polyakov loop takes a finite value at high $T$.

We find that the Polyakov loop reduces
the chiral transition temperature
and the obtained transition temperatures roughly explain
the MC results in the region $\beta \lesssim 4$.
The chiral and deconfinement transitions are found to take place
at similar temperatures. 
The W- and H-methods
exhibit qualitatively the same results,
while the Polyakov loop effects are found to be weaker in the W-method.
This is because the temporal link integral in the W-method
favors the color-singlet states
and leads to the suppression of the quark excitation.

As future works, 
we should investigate the QCD phase diagram in finite $T$ and $\mu$ 
on the basis of the effective potentials derived in this work.
Results in this direction are discussed in part in Ref.~\cite{Lat10-Miura}.

\section*{Acknowlegements}
We would like to thank Owe Philipsen and Yoshimasa Hidaka for useful discussion.
This work was supported in part
by Grants-in-Aid for Scientific Research from MEXT and JSPS
(Nos. 22-3314),
the Yukawa International Program for Quark-hadron Sciences (YIPQS),
and by Grants-in-Aid for the global COE program
`The Next Generation of Physics, Spun from Universality and Emergence'
from MEXT.

\end{document}